\documentclass[a4paper,envcountsect]{llncs}
\usepackage[utf8]{inputenc}
\usepackage[hidelinks]{hyperref}

\usepackage{underscore}
\usepackage{mathtools}
\usepackage{amssymb}
 \usepackage{wrapfig}
\usepackage{color}
\usepackage[caption=false]{subfig}
\usepackage{todonotes}
\usepackage{times}
\usepackage{microtype}
\usepackage{pgfplots}
\usetikzlibrary{calc}
\usepgfplotslibrary{colorbrewer}
\pgfplotsset{cycle list/Dark2}
\pgfplotsset{every axis plot/.append style=thick}

\DeclareMathOperator{\Prr}{P}
\DeclareMathOperator{\Runs}{Runs}
\DeclareMathOperator{\Hist}{Hist}
\DeclareMathOperator{\Cones}{Cones}
\DeclareMathOperator{\Cone}{Cone}

\DeclareMathOperator{\Val}{Val}
\DeclareMathOperator{\Exp}{\mathbb{E}}
\DeclareMathOperator{\last}{last}
\DeclareMathOperator{\first}{first}

\DeclareMathOperator{\Dist}{\mathcal{D}}
\DeclareMathOperator{\Pp}{\mathbb{P}}
\DeclareMathOperator{\Reals}{\mathbb{R}}

\newcommand{\qedd}{\hfill\ensuremath{\diamond}}

\title{Parameter-Independent Strategies \\ for pMDPs via POMDPs\protect\footnote{Extended version of a QEST 2018 paper.}}
\author{Sebastian Arming\inst{1}
\and
Ezio Bartocci\inst{2}
\and
Krishnendu Chatterjee\inst{3}
\and \\
Joost-Pieter~Katoen\inst{4}
\and
Ana Sokolova\inst{1}
}
\institute{%
University of Salzburg, Austria
\and
TU Wien, Austria
\and
IST Austria
\and
RWTH Aachen University, Germany}

\authorrunning{S. Arming, K. Chatterjee, J. Katoen, E. Bartocci, and A. Sokolova}

\begin{document}
\maketitle

\begin{abstract}
Markov Decision Processes (MDPs) are a popular class of models 
suitable for solving control decision problems in probabilistic reactive systems. 
We consider parametric MDPs (pMDPs) that include parameters in some of the 
transition probabilities to account for stochastic uncertainties of the environment 
such as noise or input disturbances.

We study pMDPs with reachability objectives where the parameter values are 
unknown and impossible to measure directly during execution, but there is a 
probability distribution known over the parameter values. We study for the first 
time computing parameter-independent strategies that are expectation optimal, 
i.e., optimize the expected reachability probability under the probability distribution 
over the parameters. We present an encoding of our problem to partially observable 
MDPs (POMDPs), i.e., a reduction of our problem to computing optimal strategies 
in POMDPs.

We evaluate our method experimentally on several benchmarks: a motivating (repeated) \emph{learner model}; 
a series of 
benchmarks of varying configurations of a robot moving on a grid; and a consensus protocol.
\end{abstract}

{\let\thefootnote\relax\footnotetext{
\scriptsize {This work was supported by  the Austrian FWF (National Research 
Network RiSE/SHiNE S11405-N23, S11407-N23  and S11411-N23) and by the 
RTG 2236 UnRAVeL funded by the German Research Foundation.}}}

\section{Introduction}\label{sec:intro}

Markov decision processes (MDPs)~\cite{Baier2008} are a popular   
class of models suitable for solving decision making 
and dependability problems in a randomized environment.  
An MDP is a state-based model representing a probabilistic process 
that satisfies the Markov property (\emph{memorylessness}), where 
for each state it is possible to choose nondeterministically 
some action-labeled transitions governing the probability distribution 
to end up in the next state.
MDPs are employed in several applications including the analysis of queueing 
systems~\cite{Sennott1998}, bird flocking~\cite{LukinaEHBYTSG17}, 
confidentiality~\cite{BaldiBCCSSS17} and robotics~\cite{Ayala2012}.

One of the main problems of interest for MDPs is the synthesis of an optimal policy 
(scheduler, strategy) choosing the sequence of 
actions that maximizes/minimizes 
the probability or the expected accumulated reward/cost 
to reach a target state.
Model-checking tools such as PRISM~\cite{KwiatkowskaNP11} 
or Storm~\cite{DehnertJK017} provide a push-button technology 
to automate such analyses and to derive simple deterministic 
and memoryless schedulers.

We study here \emph{parametric} Markov decision processes 
(pMDPs)~\cite{ChenHHKQ013,HahnHZ11}, in which (some of) 
the transition probabilities depend on a set of parameters. 
This class allows to include unknown quantities in the model 
such as the fault rate or the input disturbances that are
responsible for stochastic uncertainty. These quantities 
are often unavailable at the design time or impossible to  
measure directly at runtime.  Intuitively, a pMDP 
represents a family of MDPs---one 
for each possible valuation of the parameters.

In the past years, there has been a great effort 
to solve reachability analysis in pMDPs using 
symbolic approaches~\cite{ChenHHKQ013,HahnHZ11,QuatmannD0JK16,Cubuktepe2017,ArmingBS17,PolgreenWHA17}. 
These methods generally partition the parameter 
space in regions, associating each region to 
the optimal memoryless scheduler that 
maximizes/minimizes 
the probability to reach the target state. 
The common assumption of all these approaches 
is the possibility to observe the unknown 
quantities at some point and then to choose accordingly the best
scheduler. However, this is not always feasible.\\

\vspace*{-2.5mm}\noindent \emph{Our contribution. }
We analyze pMDPs with reachability objectives\footnote{We can deal with  objectives beyond reachability as long as they are induced by a reward structure (for applicability of the available tools), see Sections~\ref{sec:models} and~\ref{sec:encoding}.} without assuming that parameter values 
are accessible directly during execution. Specifically, \emph{we find 
parameter-independent strategies that are expectation $\varepsilon$-optimal~\cite{ArmingBS17}, for $\varepsilon \ge 0$, 
i.e., optimize the expected reachability probability given a probability distribution 
over the parameters}. 
To achieve this goal, we consider 
partially observable Markov decision processes (POMDPs) where 
an agent cannot directly observe the environment's state and 
must take decisions according to its belief on the current  
state; the belief can be updated by interacting with the environment. 

We provide an encoding of a pMDP as a POMDP where
the states consist of pairs of the original pMDP states 
and parameter values and transitions can 
occur only between states with the same parameter values.
We prove that solving the induced POMDP corresponds 
to finding parameter-independent expectation $\varepsilon$-optimal policies 
for the pMDP. Note that here memoryless policies are not sufficient for 
optimality, see the discussion on the motivating learner model in 
Section~\ref{sec:experiments}. 
We leverage algorithms such 
as \emph{point-based value iteration} (PBVI)~\cite{PineauGT03} and 
\emph{Incremental Pruning} (IP)~\cite{Cassandra:1997vz} to find 
the solution. We have implemented our approach using Storm~\cite{DehnertJK017} and AI-Toolbox~\cite{aitoolbox}. 

Finally, we evaluate our approach experimentally on several benchmarks: a motivating (repeated) \emph{learner model}; 
a series of 
benchmarks of varying configurations of a robot moving on a grid; and a consensus protocol model.\\

\vspace*{-2.5mm}\noindent \emph{Paper organization.}
Section~\ref{sec:related} discusses related work. 
In Section~\ref{sec:models} we introduce 
MDPs, pMDPs, and POMDPs. 
Section~\ref{sec:encoding} presents the encoding of a
pMDP in POMDP 
and 
the reduction result.
In Section~\ref{sec:experiments} we illustrate our approach on several 
case studies and report on experimental results, while in 
Section~\ref{sec:conclusion} we wrap up with conclusions and discussion of future work.

\section{Related Work}\label{sec:related}

Parametric probabilistic models~\cite{DC05,LMST2007} are a special class 
of Markov models where some of the transition probabilities (or rates) 
depend on one or more parameters that are not known a-priori. 
These models are particularly useful to study systems characterized 
by stochastic uncertainty due to the impossibility to access 
certain quantities (e.g., fault rates, packet loss ratios, etc.).

In the last decade, 
there was a great effort to study the problem 
of symbolic model checking of parametric probabilistic Markov chains~\cite{DC05,HahnHZ11b,BartocciGKRS11,JansenCVWAKB14,PathakAJTK15,DehnertJJCVBKA15,QuatmannD0JK16}. 
In~\cite{DC05}, Daws introduced a method to express 
 the probability to reach the  target state as a multivariate rational function
 with the domain in the parameter space.
This approach was then efficiently implemented in the PARAM1 and PARAM2 
tools~\cite{HahnHZ11b}  and included later on 
in the PRISM model checker~\cite{KwiatkowskaNP11}. 

The \emph{parameter synthesis} problem consists of (exploiting  
the generated multivariate rational function and) finding the parameter values 
that would maximize or minimize the probability to reach the target state.

\emph{Repairing} a probabilistic model~\cite{BartocciGKRS11} 
consists instead of solving a constrained nonlinear optimization problem 
where the objective function represents the minimal change 
in the transition probabilities such that the probability 
to reach the target state is constrained to a given bound.

The price to pay for these techniques is the increasing complexity 
of the multivariate rational functions in the presence of large models~\cite{Kreinovich1998bi,LMST2007},  
causing the parameter synthesis  to be also very computationally expensive.
However, the introduction of new efficient heuristics~\cite{JansenCVWAKB14,PathakAJTK15,DehnertJJCVBKA15,QuatmannD0JK16} 
has helped to alleviate this problem by supporting 
the parameter synthesis for quite large models.

 This symbolic approach to parameter synthesis has been recently 
 extended to handle also the nondeterministic choice in 
 parametric Markov decision processes (pMDPs)~\cite{ChenHHKQ013,HahnHZ11,QuatmannD0JK16,Cubuktepe2017,ArmingBS17,PolgreenWHA17},
 where each different sequence of inputs can induce a
 distinct Markov chain, resulting potentially in several  
multivariate rational functions.

The parameter synthesis problem for pMDP 
consists in solving a nonlinear program (NLP) with multiple objectives.
Recently, the authors in~\cite{Cubuktepe2017} have shown 
that many NLPs related to pMDP belong to
a certain class of nonconvex optimization problems
called signomial programs (SGPs). In the same paper, they have also  introduced 
an approach to relax nonconvex constraints in SGPs generating 
geometric programs, a particular class of convex programs 
that can be solved in a number of steps that is polynomial in 
the number of variables.

Another approach proposed in~\cite{ChenHHKQ013} is based on sampling techniques
(i.e.,  Metropolis-Hastings algorithm,
particle swarm optimization, and cross-entropy method) that are used 
to search the parameter space.  These heuristics usually do not guarantee 
that global optimal parameters will be found. Furthermore, when the regions 
of the parameters satisfying a requirement are very small, a large number of 
simulations is required.

All the proposed methods provide a map that relates the regions of the parameter 
space to the optimal memoryless scheduler that maximizes/minimizes 
the probability to reach the target state.  The underlying assumption 
for these approaches is the possibility to measure the parameters during system execution.  Once
the values of the parameters have been measured, one can use this map to choose 
the best policy.  In this paper, we consider a different assumption
with respect to the state of the art: \emph{We want to synthesize an expectation 
$\varepsilon$-optimal scheduler for pMDP that is independent from the possibility to 
measure the parameters.} To achieve this goal we show how to 
recast the problem into finding an $\varepsilon$-optimal policy for a partially observable 
Markov decision process (POMDP) after providing a suitable encoding. While we are not 
aware of any other work that establishes such a correspondence, it is 
worth mentioning that instead parameter synthesis for parametric Markov chains  has 
been recently employed to find permissive finite-state controllers 
for POMDPs in~\cite{abs-1710-10294}.

The qualitative analysis of POMDPs has been widely studied: complexity results have been established~\cite{CDH10a,BGB12} and symbolic algorithms~\cite{CCD16}. However, for the general quantitative problem and its approximation the computational questions are undecidable~\cite{MHC03,DBLP:journals/ai/ChatterjeeC15}. Despite the undecidability there are several practical approaches such as point-based methods~\cite{Spaan:2011cs}.

\section{The Models}\label{sec:models}

In this section we introduce the models of importance for this paper: MDPs, parametric MDPs, and partially observable MDPs. The models can be arbitrarily large, we do not impose restrictions on the size for the theoretical part of the paper.

\subsection{Markov Decision Processes -- MDPs}\label{sec:mdp}

A (discrete) probability distribution on a set $S$ is a function $\mu\colon S \to [0,1]$ with the property $\sum_{s \in S}\mu(s) = 1$. 
By $\Dist{S}$ we denote the set of all (discrete) probability distributions on  $S$. For $s \in S$, we write $\delta_s$ for the Dirac distribution that assigns $1$ to $s$. 

\begin{definition}[Markov Decision Process -- MDP]\label{def:mdp}
A Markov Decision Process (MDP) is a tuple
$M = (S, A, T, i)$
where:
\begin{itemize}
\item $S$ is a set of states,
\item $A$ is a set of actions,
\item $T\colon S \times A \to \Dist{S}$ is the transition function, and
\item $i \in \Dist{S}$ is the initial state distribution. \qedd
\end{itemize}
\end{definition}
From a given state with a given label, an MDP makes a step to a probability 
distribution over states that describes the probability of reaching a next state.  
As usual, we write $s \stackrel{a}{\to} \mu$ for $\mu = T(s,a)$.  We will 
write $s \stackrel{a,p}{\to} t$ for $s \stackrel{a}{\to} \mu$ and $p = \mu(t)$, 
as well as $s \stackrel{a}{\to} t$ for $s \stackrel{a,p}{\to} t$ with $p > 0$. 
It is also common to write $T(t|s,a)$ for $T(s,a)(t)$.

\begin{remark} In our definition no action is disabled in any state. This is somewhat unusual for MDPs, but very common for partially observable MDPs which we are interested in.
\end{remark}

 \noindent   A \emph{run} (also called path or play) of an MDP is an infinite sequence $s_0,a_0,s_1,a_1,\dots$ in $(S \times A)^\omega$ of
    states and actions such that $s_i \stackrel{a_i}{\to} s_{i+1}$ for all $i\geq 0$. When convenient, we will also write $s_0\cdot a_0\cdot s_1\cdot a_1\dots$ for the run $s_0,a_0,s_1,a_1,\dots$.
    A \emph{history} $h$ is a finite prefix of a run in $(S\times A)^* \times S$. We write $\first(h)$ and $\last(h)$ for the first and last state in a history $h$, respectively.
    The \emph{cone} $\Cone(h)$ of a history $h$ is the set of all runs with prefix $h$.
    We write $\Runs$ for the set of all runs, $\Hist$ for the sets of all
    histories, and $\Cones$ for the smallest $\sigma$-algebra containing the cones of all histories. %

\begin{definition}[MDP policy]\label{def:mdp-policy}
    A policy (strategy, scheduler) $\pi$ for an MDP $M$ is a map
    \[\pi \colon \Hist \to \Dist{A}\]
    from histories to probability distributions over the actions.
    It is a deterministic policy if the image of $\pi$ consists only of Dirac distributions.
    It is a memoryless (or \emph{Markov}) policy if $\pi(w \cdot s) = \pi(s)$ for $w \in (S\times A)^*$ and  $s \in S$.\qedd
\end{definition}

A policy $\pi$ together with the initial state distribution $i\in \Dist S$ induces a probability
space $(\Runs,\Cones,\Pp_{\pi,i})$, i.e., a probability measure $\Pp_{\pi,i}$ on the measurable space $(\Runs,\Cones)$. 
This construction is done in several steps: First, for a given state $s$, we consider a function $\Prr_{\pi,s}$ assigning a number in $[0,1]$ to cones of histories in $\Hist$. It is defined inductively as follows.
We set $\Prr_{\pi,s}(\Cone(h)) = 1$ if $h = s$ and $\Prr_{\pi,s}(\Cone(h)) = 0$ if $h = t \neq s$. For $h = w \cdot a \cdot t$ we set
 $\Prr_{\pi,s}(\Cone(h)) = \Prr_{\pi,s}(\Cone(w)) \cdot \pi(w)(a)\cdot T(\last(w), a)(t)$. 
By Carathéodory's extension theorem, the function $\Prr_{\pi,s}$ extends to a unique measure on the measurable space $(\Runs, \Cones)$ which we denote by $\Pp_{\pi,s}$.
Finally, given the initial state distribution $i\in \Dist S$, $\Pp_{\pi,i}$ is the measure on $(\Runs, \Cones)$ defined as
$\Pp_{\pi,i} = \sum_{s \in S} i(s) \cdot \Pp_{\pi,s}.$

We write $\Exp_{\pi,i}$ for the expectation operator of $\Pp_{\pi,i}$. 
Recall that the expectation operator of a measure $\mu$ on a measurable space 
$(X, \Sigma)$ is defined as $E_\mu(f) = \int f d\mu$ for a measurable function 
$f\colon X \to \Reals$ where we consider the Borel $\sigma$-algebra on the reals.
Hence $\Exp_{\pi,i}(f) = \int f d\Pp_{\pi,i}$ for a measurable function $f\colon \Runs \to \Reals$. We may sometimes decorate the notation of the measures and the expectation operators by superscript $M$, to emphasize the involved model. 

We can now specify what it means to solve an MDP.
\begin{definition}[Objective, value, solution of MDP]\label{def:solution-mdp}
Given an MDP $M = (S, A, T, i)$, a \emph{Borel objective}, also called return, is  
a measurable function $r\colon \Runs \to \Reals$. 
The \emph{value} of the MDP $M$ for the objective $r$ is defined as $\Val(r) = \sup_\pi \Exp_{\pi,i}(r)$.  
A \emph{solution} to an MDP $M$ regarding the objective $r$ is a policy $\pi$
with $\Exp_{\pi,i}(r) =
    \Val(r)$.  A policy $\pi$ is an \emph{$\varepsilon$-solution}, for $\varepsilon > 0$, of $M$ with respect to $r$ if $\Exp_{\pi,i}(r)$ is $\varepsilon$-close to $\Val(r)$. \qedd
\end{definition}

Note that a solution to an MDP need not exist.
We will say objective for a Borel objective.
Some objectives arise via the \emph{payoff} or \emph{accumulated reward} 
of runs. For MDPs with such reward-based objectives, a solution always exists. Solving (partially observable) MDPs often refers to solving reward-based objectives.  

\begin{definition}[MDP with rewards]\label{def:rewards}
An MDP with rewards is a tuple $M_R = (M,R)$ where $M = (S, A, T,i)$ is an MDP and $R\colon S \times A \times S \to \mathbb{R}$ is the reward function.\qedd
\end{definition}

Upon performing a transition, an MDP with rewards collects reward as described by the reward function. We will sometimes write $s \stackrel{a,p,r}{\to} t$ for $s \stackrel{a,p}{\to} t$ and $R(s,a,t) = r$, and $s \stackrel{a,r}{\to} t$ for $s \stackrel{a}{\to} t$ and $R(s,a,t) = r$. If clear from the context, we will drop the subscript $R$ in an MDP $M_R$ with rewards.

Reward structures may induce objectives as follows:
The (undiscounted) accumulated reward of a run in an MDP with rewards is 
\begin{equation}\label{eq:acc-reward-obj}
	r_R(s_0,a_0,s_1, a_1 \dots) = \sum_{i \ge 0}  R(s_{i}, a_{i}, s_{i+1}).
\end{equation}
The accumulated reward induces the \emph{reward objective} $r_R$ if the above assignment  defines a measurable function $r_R \colon \Runs \to \mathbb{R}$.\footnote{The discounted accumulated reward objective is defined in a similar way, by adding a factor $\gamma^i$ to the $i$-th summand in~(\ref{eq:acc-reward-obj}) with $\gamma \in [0,1)$ being the discount factor. For solving reachability objectives, undiscounted rewards are sufficient.}

\subsection*{Reachability Objectives}
Of special interest to us is optimizing \emph{reachability}, that is optimizing the
probability to reach a target state (or a set of target states).
Computing extremal (i.e., maximal/minimal) reachability probabilities is
at the heart of MDP model checking: PCTL and LTL model checking boil down to
computing reachability probabilities. The same holds for omega-regular
properties: determining the maximal probability of any omega-regular
property $\varphi$ in an MDP $M$ amounts to determining the maximal
probability to reach an accepting end component in the product of $M$
with a deterministic omega-automaton for $\varphi$. Verifying PCTL
properties under fair policies, i.e., policies that can almost surely
reach a state satisfying some fairness constraint, can also be reduced
to computing reachability probabilities.

\begin{definition}[Reachability objective]\label{def: reach-obj}
Let $M = (S, A, T, i)$ be an MDP and $t \in T$ a state. The \emph{reachability objective} 
$r_t$ of reaching the state $t$ is given by the indicator function of the set $\Runs_t$ of runs that reach $t$, i.e.,
$$\Runs_t = \{s_0,a_0,s_1,a_1, \dots  \in \Runs \mid \exists i \ge 0. s_i = t\}$$
and $r_t(\rho) = 1$ if  $\rho \in \Runs_t$ and $r_t(\rho) = 0$ otherwise. \qedd
\end{definition}

Note that $\Runs_t$ is a measurable set, i.e., $\Runs_t \in \Cones$ and hence $r_t$ is a measurable function. Moreover, a solution to the reachability objective $r_t$ is a policy $\pi$ that maximizes the reachability probability, as $\Exp_{\pi,i}(r_t) = \Pp_{\pi,i}(\Runs_t)$.

Reachability objectives are induced by reward structures as follows.
Given an MDP $M=(S,A,T,i)$ and a target state $t\in S$ we construct the MDP with rewards
$M_t=(S,A,T_t,i, R_t)$ where
    $$T_t(s,a)=\begin{cases}
        \delta_t & \text{if } s=t\\
        T(s,a) & \text{otherwise}
    \end{cases}  \qquad 
    R_t(s,a,s')=\begin{cases}
        1 &\text{if } s\neq t \text{ and } s'=t\\
        0 &\text{otherwise .}
    \end{cases}$$

The following standard property relates the accumulated reward of $M_t$ to solving the reachability objective and is not difficult to show. (The condition $i(t) = 0$ is technical: in $M_t$ the history $t$ does not accumulate any reward.)

\begin{proposition}
    For any policy $\pi$ for an MDP $M = (S,A,T,i)$ with $t \in S$ and $i(t)=0$, the
    probability to reach $t$ in $M$ under $\pi$ is the
    (undiscounted) accumulated reward of $\pi$ in $M_t$, i.e., $\Pp_{\pi,i}^M(\Runs_t) = \Exp_{\pi,i}^{M_t}(r_{R_t})$. \qed
\end{proposition} 

The finite-horizon reachability objective is reachability within $k$ steps for $k \in \mathbb{N}$ being called \emph{horizon}. This objective is induced by the rewards via 
\begin{equation}\label{eq:fin-hor}
r_{R_t}^k(s_0,a_0,s_1, a_1 \dots) = \sum_{0 \le i \le {k-1}}  R_t(s_{i}, a_{i}, s_{i+1})
\end{equation}
with $R_t$ being the reachability reward structure in $M_t$.
\subsection{Parametric MDPs -- pMDPs}\label{sec:pmdp}

\begin{definition}[Parametric MDP -- pMDP\footnote{We use the abbreviation pMDP rather than PMDP as it is common in the recent literature, see e.g.~\cite{Cubuktepe2017,PolgreenWHA17} and as it reminds of the parameter $p$.}]\label{def:pmdp}\,
 A \emph{parametric Markov Decision Process} (pMDP) is a tuple 
$M = (S, A, X, T,i)$
where $S$, $A$, and $i$ are as in the definition of an MDP and
\begin{itemize}
\item
$X$ is the parameter space,
\item 
$T\colon S \times A \to (\Dist S)^X$ is the transition function.\qedd%
\end{itemize}
\end{definition}

For a pMDP $M$ as above and a parameter point $x\in X$ we write $M(x)$ for the
\emph{evaluation of $M$ at $x$}, that is the MDP $M(x)=(S,A,T_x,i)$ with
$T_x(s,a)=T(s,a)(x)$.

We are interested in finding $\varepsilon$-optimal policies independent of the parameter,
i.e., policies that are somehow $\varepsilon$-optimal over the whole parameter space $X$.
Since for pMDPs the expected return is a function in the parameter, it is
not a priori clear what optimality criterion to choose ---
see~\cite{ArmingBS17} for several alternatives.
Here we consider the case where a parameter distribution $p\in \Dist X$ is
given and we optimize the expected reward given $p$ (this setting is called
\emph{expectation optimal} in~\cite{ArmingBS17})---as formalized below.

Runs, histories, and policies are defined similarly as for MDPs including in addition the parameter value, and the probability measure now also depends on $p$. For a parameter space $X$, the sample space is
$\Runs_X =  \{(x,\rho) \mid x\in X, \rho \in \Runs(M(x))\}$;  
the set of histories is
$\Hist_X =  \{(x,h) \mid x\in X, h \in \Hist(M(x))\}$; 
and the $\sigma$-algebra is the smallest $\sigma$-algebra $\Cones_X$ that contains the sets
\[ \{\{x\}\times\Cone(h) \mid x\in X, h \in \Hist(M(x))\}.\]

\begin{definition}[pMDP policy]\label{def:pMDP-policy}
A \emph{policy} $\pi_X$ of a pMDP $M$ is a map $\pi_X \colon \Hist_X \to \Dist{A}$ that is \emph{independent from the parameters}, i.e., that satisfies the property
\begin{equation}\label{eq:policy-req-pmdp}
	\pi_X(x,h) = \pi_X(y,h) \quad \text{ for all } x, y \in X, h \in \Hist(M(x)) \cap \Hist(M(y)).
\end{equation}
	
\end{definition}
Note that in general $\Hist(M(x))$ may differ from $\Hist(M(y))$ for $x \neq y$, as different parameter values may make the transition probabilities of certain transitions equal to zero. Furthermore, note that a pMDP policy, due to the requirement~(\ref{eq:policy-req-pmdp}) can equivalently be defined as a map $\pi_X\colon \bigcup_{x \in X} \Hist(M(x)) \to \Dist{A}$.

A pMDP policy $\pi_X$, induces a family of MDP policies $(\pi_x \mid x \in X)$ with $\pi_x$ a policy for $M(x)$ by projection, i.e., $\pi_x(h) = \pi_X(x,h)$ for all $h \in \Hist(M(x))$. 

The measurable space $(\Runs_X, \Cones_X)$ is (isomorphic to) the disjoint union (coproduct) measurable space 
$$\Runs_X = \coprod_{x \in X} \Runs(M(x)), \quad \Cones_X = \{\coprod_{x \in X} A_x \mid A_x \in \Cones(M(x))\}$$
and by $\Pp_{\pi_X,i,p}$ we denote the measure that is the $p$-convex combination of the measures $\Pp_{\pi_x,i}^{M(x)}$, i.e.,
\begin{equation}\label{eq:pmdp-measure}
\Pp_{\pi_X,i,p}(\coprod_{x \in X} A_x) = \sum_{x \in X}p(x)\cdot\Pp_{\pi_x,i}^{M(x)}
(A_x).
\end{equation}

\begin{remark} \label{rem-ext-pMDP-measure}
Note that $\Pp_{\pi_X,i,p}$ is the unique extension of the assignment
$$\Prr_{\pi_X,i,p}(\{x\}\times\Cone(h))=p(x) \cdot \Pp{}^{M(x)}_{\pi_x,i}(\Cone(h))$$
to the measurable space $(\Runs_X, \Cones_X)$.	
\end{remark}

We write $\Exp_{\pi_X,i,p}$ for the expectation operator of $\Pp_{\pi_X,i,p}$. The \emph{value} of a pMDP $M$ given parameter distribution $p$ and objective $r$ is
$\Val(p,r) = \sup_{\pi_X} \Exp_{\pi_X,i,p}(r)$ where the supremum is taken over all pMDP policies $\pi_X$.

\begin{definition}[Expectation $\varepsilon$--optimal policy]\label{def:exp-opt-policy}\,  A policy $\pi_X$ is \emph{expectation $\varepsilon$-optimal} for a pMDP $M$ iff $\Exp_{\pi_X,i,p}(r)$ is $\varepsilon$-close to $\Val(p,r)$. \qedd
\end{definition}

\subsection{Partially Observable MDPs -- POMDPs}\label{sec:pomdp}
\begin{definition}[Partially observable MDP -- POMDP]\label{def:pomdp}
A \emph{partially observable MDP} (POMDP) is a tuple 
$M = (S, A, T, i, \Omega, O)$
where:
\begin{itemize}
\item
$(S,A,T, i)$ is the underlying MDP,
\item
$\Omega$ is the set of observations,
\item 
$O\colon S \to \Omega$ is the observation function.\qedd%
\end{itemize}
\end{definition}

Note that our observation function is deterministic and only state-dependent, 
which is not a restriction~\cite{ChatterjeeCGK16}.
Runs and histories of a POMDP are the runs and the histories of its underlying 
MDP. The reward structure for a POMDP is a reward structure of its underlying MDP.
The observation function $O$ extends naturally to runs and histories as follows, 
by slight abuse of the notation we denote all these functions by $O$. 
We have $O\colon \Runs \to (\Omega\times A)^\omega$  given by
$$O(s_0, a_0, s_1, a_1, \dots)\,\, = \,\, O(s_0), a_0, O(s_1), a_1, \dots$$
and similarly we define $O\colon \Hist \to (\Omega\times A)^*\times\Omega$. 

The crucial difference to MDPs is that the policy of a POMDP can only observe
the observations but not the states directly:
\begin{definition}[POMDP policy]\label{def:pomdp-policy}
    A \emph{policy} $\pi$ for a POMDP $M$ is a policy for the underlying MDP of $M$ 
    with the additional requirement that $\pi(h)=\pi(h')$
    whenever $O(h)=O(h')$, for all $h,h'\in \Hist$.\qedd%
\end{definition}

POMDP policies, together with the initial state distribution, also induce a probability measure over runs. The measurable space is again $(\Runs, \Cones)$. The measure $\Pp_{\pi,i}$ is defined in exactly the same way as for the underlying MDP, and $\Exp_{\pi,i}$ again denotes the expectation operator. 
The value of a POMDP $M$ on an objective $r$, also denoted by  $\Val(r)$, is defined as $\Val(r) = \sup_\pi \Exp_{\pi,i}(r)$ where the supremum is taken over all \emph{POMDP policies} $\pi$.
A POMDP policy $\pi$ is a \emph{solution} of the POMDP $M$ for an objective $r$, iff  $ \Exp_{\pi,i}(r) = \Val(r)$; it is an $\varepsilon$\emph{-solution} for $\varepsilon >0$ iff  $ \Exp_{\pi,i}(r)$ is $\varepsilon$-close to $\Val(r)$.

Finite-horizon accumulated reward objectives as defined in~(\ref{eq:fin-hor}) are  by far the most studied class of POMDP objectives; one might even say that solving such objectives is \emph{the} POMDP problem~\cite{PineauGT03,Cassandra:1997vz}. 

\begin{remark}\label{rem:krish-POMDP}
	We note two important facts for solutions of POMDPs:
	\begin{itemize}
		\item[(1)] For POMDPs, deterministic policies are not a restriction (they are as powerful as randomized policies, but can require more memory) for any Borel objective, see~\cite[Lemma~1,Theorem~7]{Krish:Randomness-for-free}.
		\item[(2)] For POMDPs with reachability objectives, for $\varepsilon$-approximation with $\varepsilon>0$, finite-memory policies are sufficient for optimality. This is because given any $\varepsilon>0$, there exists a finite horizon $N_\varepsilon$, such that reachability within $N_{\varepsilon}$ steps $\varepsilon$-approximates the optimal reachability probability, and for finite-horizon reachability optimal finite-memory policies are sufficient. 
	\end{itemize}
\end{remark}

\section{The Encoding}\label{sec:encoding}

In this section we reduce the problem of finding an expectation-optimal policy for a pMDP to the problem of solving a POMDP, by presenting an encoding of pMDPs to POMDPs that will relate the policies in the desired way. 

The main technical observation of our paper, the observation that enables the method of finding parameter-independent optimal policies for pMDPs via solving the induced POMDP, is the encoding and the correspondence result, Theorem~\ref{thm:correspondence} below. We start with presenting the encoding.

\begin{definition}[Induced POMDP]\label{def:encoding}
Given a pMDP $M=(S,A,X,T,i)$, its \emph{induced POMDP} is 
$M'=(S\times X, A, T', S, O)$
with:
    $$T'((s,x),a)(s',x') = T(s,a)(x)(s') \cdot \delta_x(x')$$
    and 
    $ O((s,x)) = s$. \qedd
\end{definition}

Hence, the encoding, i.e., the induced POMDP of a pMDP $M$ is a POMDP with much larger state space: new states are pairs of states of $M$  (``old" states) and parameter values in $X$. Transitions are only possible among new states with the same parameter value, i.e., transitions can not change the parameter values, and the transitions are inherited from the pMDP. Observations are the old states, i.e., in a new state $(s,x)$ we can observe the old state $s$ but not the parameter value $x$.

Our correspondence result is a consequence of the classical ``change of variable" result of measure theory, which we recall next.

\begin{theorem}[{\cite[Theorem~VIII.C]{Halmos:MT-Book}}] \label{thm:halmos}
Let $(X, \Sigma)$ and $(X', \Sigma')$ be measurable spaces, $f\colon (X,\Sigma) \to (X',\Sigma')$ a measurable function, $\mu\colon \Sigma\to \mathbb R^+$ a measure, and $\varphi' \colon X' \to \mathbb{R}^{+} \cup\{ \infty\}$ a measurable function. 

Let $\mu' \colon X' \to \mathbb R^+$ be the push-forward measure of $\mu$ along $f$, i.e. $\mu' = \mu \circ f^{-1}$ and let $\varphi = \varphi' \circ f \colon X \to \mathbb{R}^{+} \cup\{ \infty\}$. 

Then $$\int_X \varphi\, d\mu = \int_{X'} \varphi' \,d\mu',$$
that is, if one of the integrals exists, the other does too and they are equal.  \qed
\end{theorem}

At this point, let us denote by $\Runs'$, $\Hist'$ and $\Cones'$ the runs, histories, and the $\sigma$-algebra generated by the cones of histories of the encoding $M'$. Note that a policy of $M'$ is a map $\pi' \colon \Hist' \to \Dist{A}$. Moreover, note that 
\begin{eqnarray*}\Runs' &=& \{(s_0,x),a_0, (s_1,x),a_1, \dots \mid x \in X \text{ and } s_0,a_0,s_1,a_0, \dots \in \Runs(M(x)) \}\end{eqnarray*}
and analogously for histories. This is a consequence of the construction of the encoding, i.e., of the fact that every run of $M'$
    involves a single parameter point as $T'((s,x),a)(s',x')=0$ whenever $x\neq x'$.

We now state several direct consequences of the definitions needed for the correspondence result. The proofs of all results are in Appendix~\ref{app:proofs}.

\begin{lemma}\label{lem:runs-1-1}
    The function $f: \Runs_X \to \Runs'$  defined by 
    \[
        f(x,s_0\cdot a_0\cdot s_1\cdots) = (s_0,x)\cdot a_0\cdot (s_1,x)\cdots
    \]
    is an isomorphism between the measurable spaces $(\Runs_X, \Cones_X)$ of the pMDP $M$ and $(\Runs', \Cones')$ of its induced POMDP $M'$. \qed
\end{lemma}

\noindent This means that $f$ is a bijection and both $f$ and $f^{-1}$ are measurable functions. In the sequel, we also write $f$ for the bijection from $\Hist_X$ to $\Hist'$ defined in the same way.
Note that $f$ maps generators of $\Cones_X$ to generators of $\Cones'$ as $$f(\{x\}\times \Cone(h_x)) = \Cone(f(x,h_x))$$ for any $x \in X$ and $h_x \in \Hist(M(x))$. 
Moreover, to state the obvious, $f$ is related to the observation map $O$ as follows: $O(f(x,h)) = h$.

\begin{lemma}\label{lem:policy-1-1}
There is a bijective correspondence $\Phi$ between the policies of the pMDP $M$ and the policies of its induced POMDP $M'$ given by
$\Phi(\pi_X) = \pi_X \circ f^{-1}$. Its inverse acts as $\Phi^{-1}(\pi') = \pi' \circ f$. \qed
\end{lemma}

\noindent We are now able to relate the induced measures in a pMDP and its encoding.

\begin{lemma}\label{lem:measure-eq}
    Given a pMDP $M=(S,A,X,T,i)$,
    parameter distribution $p$,
    and policy $\pi_X$,
    we have
    \[
        \Pp{}^{M}_{\pi_X,i,p}=\Pp{}^{M'}_{\pi',i'}\circ f
\]
    where $M'$ is the induced POMDP of $M$, $\pi'= \Phi(\pi_X)$, and $i'(s,x)=i(s)\cdot p(x)$. \qed
\end{lemma}

\noindent Now the correctness of the encoding follows easily from the next result.

\begin{theorem}\label{thm:correspondence}
    Given a pMDP $M=(S,A,X,T,i)$, parameter
    distribution $p \in \Dist{X}$, policy $\pi_X$, and an objective function $r$, we have
    \begin{equation}\label{eq:corr-thm}
        \Exp^M_{\pi_X,i,p}(r) = \Exp^{M'}_{\pi',i'}(r')
    \end{equation}
    where $M'$ is the induced POMDP of $M$, $\pi'= \Phi(\pi_X)$, $r' = r \circ f^{-1}$, and $i'(s,x)=i(s)\cdot p(x)$.
    The opposite also holds: Given a policy $\pi'$ of the induced POMDP $M'$ of $M$, Eq.~(\ref{eq:corr-thm}) holds for $\pi_X = \Phi^{-1}(\pi')$. \qed
\end{theorem}

As a consequence, $\Val^M(p,r) = \Val^{M'}(r')$ and the policy $\pi_X$ is expectation $\varepsilon$-optimal for $M$ if and only if $\pi'= \Phi(\pi_X)$ is an  $\varepsilon$-solution of $M'$, for $\varepsilon \ge 0$.

\section[Experiments]{Experiments\protect\footnote{All of our code and models, as well as detailed results of the experiments can be found at~\url{http://github.com/sarming/pMDP-Toolbox}.}}\label{sec:experiments}

\sloppypar{In this section, we present several case studies to illustrate our approach of finding an expectation $\varepsilon$-optimal policy of a pMDP using an existing POMDP solver.
The solver we use is AI-Toolbox~\cite{aitoolbox}, a well-known suite of algorithms for Markov models, which includes several algorithms for POMDPs with finite-horizon accumulated-reward objectives.
For our case studies we evaluate two algorithms: \emph{point-based value iteration} (PBVI)~\cite{PineauGT03} and \emph{Incremental Pruning} (IP)~\cite{Cassandra:1997vz}, for reasons explained in Appendix~\ref{app:AI-toolbox}.

The PBVI implementation in AI-Toolbox does not generate beliefs on the fly (as in the anytime algorithm described in~\cite{PineauGT03}), but generates a fixed number of beliefs upfront. 
First the simplex corners (i.e. the Diracs) and the midpoint are generated, then more are sampled (the point-based algorithms differ mainly in how the beliefs are sampled).
Without at least the simplex corners and midpoint, the results are quite off - we chose to always pick at least all of those.

\paragraph{Experimental Setup.} 
All the models that we analyze are described in PRISM file format.
We use Storm~\cite{DehnertJK017} to parse the files and build the parametric model, that we then translate and pass to AI-Toolbox. Both Storm and AI-Toolbox have Python bindings allowing us to perform our encoding in Python.
The experiments reported here ran on a NUMA machine with four 16-core 2.3GHz AMD Opteron 6376 processors, 504GB of main memory, and Linux kernel version 4.13.0.
We used BenchExec~\cite{Beyer:2015ip} to run experiment series.\looseness=-1

\paragraph{Selected Case Studies.} 
 We start by discussing a motivating \emph{learner model} example that shows all important aspects. Then we present some results on a model of a robot moving on a grid and a consensus protocol model. 
 
It is important to note here that 
many of the pMDP models studied (e.g., in the PARAM benchmarks \cite{PARAM:CaseStudies}) exhibit only a weak form of nondeterminism, where the optimal policy does not depend on the parameters, i.e., the optimal policy is the same for any values of the parameters. Examples of such are, e.g., the Bounded Retransmission Protocol (BRP) and  Zeroconf. Consensus is the exception, and we have solved some instances of Consensus as reported below.
In the examples used in~\cite{PolgreenWHA17}, showing how to obtain policies that optimize learning the parameter values, the optimal policy is again independent of the parameters.

\subsubsection*{Motivating Example -- the (Repeated) Learner Model} 

\begin{figure}[hbt]\centering
    \subfloat[pMDP Model]{\centering%
\resizebox{0.5\linewidth}{!}{%
\begin{tikzpicture}[auto,xscale=1.3,yscale=1.1, font=\small, node distance=100mm,>=latex]
    \tikzstyle{round}=[thick,draw=black,circle]
    \tikzstyle{choice}=[->,thick]
    \tikzstyle{prob}=[dotted,->,thick]
    \definecolor{target}{RGB}{126,164,38}
    
    \node[round] (s) at (0,0) {$s$};
    \coordinate (se) at (1,0);
    \node[round] (a) at (2,1) {$a$};
    \node[round] (b) at (2,-1) {$b$};
    \node[round] (c) at (3,0) {$c$};
    \coordinate (ca) at (3.5,0.5);
    \coordinate (cb) at (3.5,-0.5);
    \coordinate (cc) at (0.5,0.5);

    \node[round,fill=target,double] (t) at (5,1) {$t$};
    \node[round,fill=red] (x) at (5,-1) {$x$};

	\draw[choice] (-0.4,0.4) to (s);

    \draw[choice] (s) to node{$e$} (se) ;
    \draw[prob] (se) to node[left]{$p$} (a);
    \draw[prob] (se) to node[left]{$1-p$} (b);
    
    \draw[choice] (a) to node[above]{$e$} ($(a)!0.5!(c)$) ;
    \draw[choice] (b) to node[below]{$e$} ($(b)!0.5!(c)$) ;
    \draw[prob] ($(a)!0.5!(c)$) to node[above]{$1$} (c);
    \draw[prob] ($(b)!0.5!(c)$) to node[below]{$1$} (c);
    
    \draw[choice] (c) to node[above=1mm]{$a$} (ca);
    \draw[prob] (ca) to node{$p$} (t);
    \draw[prob] (ca) to node[right]{$1-p$} (x);
    
    \draw[choice] (c) to node[below=1mm]{$b$} (cb);
    \draw[prob] (cb) to node[right] {$1-p$} (t);
    \draw[prob] (cb) to node[below] {$p$} (x);
    
    \draw[choice,gray] (c) to [out=100, in=45, looseness=1.6] node[above] {$c$} (cc);
    \draw[prob,gray] (cc) to node[above] {$1$} (s);
\end{tikzpicture}}%
\label{fig:learner}} \qquad
    \subfloat[Policies]{\centering%
    \includegraphics[width=0.4\linewidth]{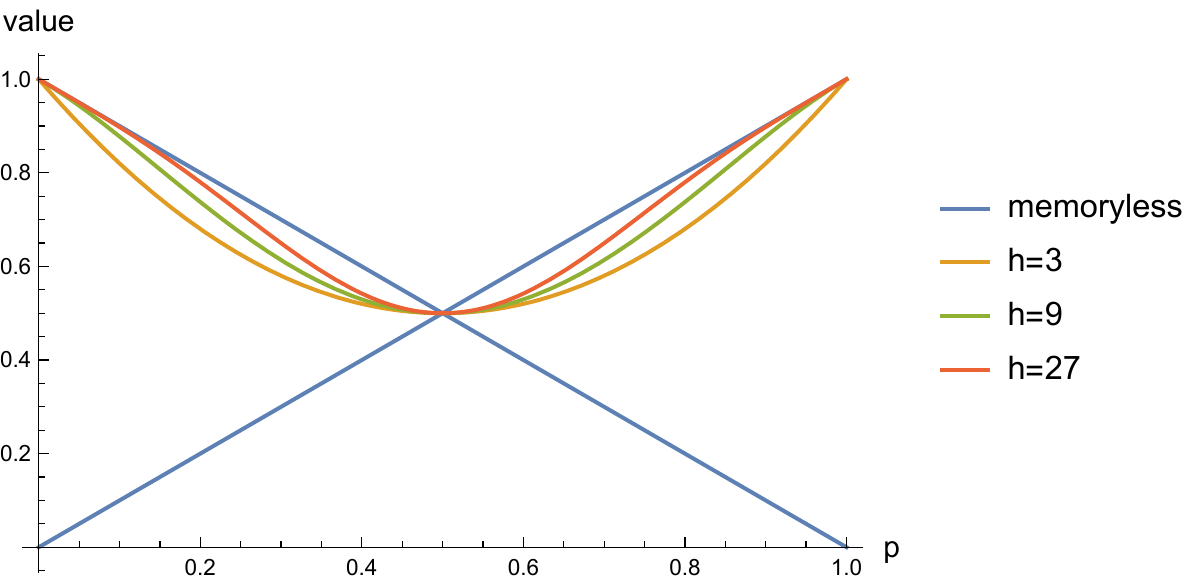}%
  \label{fig:learnerpol}}
  \caption{(Repeated) learner}
\end{figure}
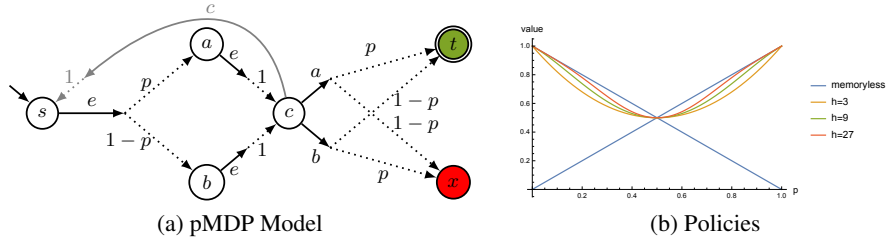
Figure~\ref{fig:learner} shows the \emph{learner model}, mentioned in~\cite{ArmingBS17}, with initial state $s$ and target state $t$, ignoring for the moment the grey (loop) transition labelled with action $c$.

After two steps we end up in state $c$, having visited  state $a$ with probability $p$ and state $b$ with probability $1-p$.
Here is the only choice in the model: between the actions $a$ and $b$.
Even though we assume $p$ to be inaccessible, we can do better than flipping a coin:
If we are given a prior belief over the parameter (an initial parameter distribution) we can use Bayesian inference to update this belief with the information that either state $a$ or $b$ was visited.
As a concrete example, assume we start with the uniform distribution over the parameter values as prior and state $a$ was visited. Doing the calculation gives a posterior distribution with higher probability that the parameter value is close to $1$ than to $0$, suggesting that $a$ is the better action to choose.

Let $\pi$ be the policy that chooses action $a$ in state $c$ when state $a$ was visited and action $b$ when state $b$ was visited.
Figure~\ref{fig:learnerpol} shows $\pi$ (labelled $h=3$) and the two memoryless policies (always $a$ and always $b$).
  The policy $\pi$ is clearly better in expectation than the two memoryless ones.

Actually, $\pi$ is the optimal policy (among the 4 possible  deterministic policies) when assuming a uniform parameter distribution.
For a concrete parameter value $x\in[0,1]$, the probability to reach $t$ under $\pi$ in the evaluated MDP $M(x)$ is $x^2+(1-x)^2$.
Putting things together, using Eq.~(\ref{eq:pmdp-measure}), we get that for a uniform parameter distribution the expected return is (a discrete approximation of) $\int_0^1 x^2 + (1-x)^2\, dx =  2/3$.

We consider the uniform distributions over 2, 3, 5, 10, 20, 50, 100, 200, 500, and 1000 evenly spaced points between 0 and 1.
The expected return depends on the distribution: for just 2 points (the distribution assigning 1/2 to 0 and 1/2 to 1) the optimal policy provides probability 1 for reaching $t$; while for 1000 points the expected return is $0.6670$.
IP can solve all mentioned instances and generates $\pi$ verifying that it is optimal.

The learner is inherently a finite-horizon model, as nothing happens after three steps. When we add the grey transition, we obtain the \emph{repeated learner model} in which we can repeat the "experiment"  getting closer and closer to the "sea surface"~\cite{ArmingBS17} given by the memoryless policies.
The larger the horizon, the more experiments the $\varepsilon$-optimal policy runs.
Only an odd number of experiments gives an actual improvement because we need a majority: having two experiments is as good as just having one. Therefore the value increases at $h=3(2n+1)$, 
see Figure~\ref{fig:learnerpol} and Table~\ref{tab:iprep}.

Figure~\ref{fig:rep} shows the results of the experimental evaluation of the repeated learner with increasing horizon and up to 100 parameter points. IP provides better policies than PBVI with respect to the same number of points in the parameter space. However, it does not scale as well as PBVI and we had to drop it for more complex models.

\begin{table}[hbt]\centering
\subfloat[$h=3$ (i.e. without repeating)]{%
  \begin{tabular}{l|cccc}
      points & states & time (s) & value & nodes \\
    \hline
2 & 14 & 0.001 & 1 & 3 \\
5 & 35 & 0.003 & 0.75 & 7 \\
10 & 70 & 0.005 & 0.703 & 7 \\
20 & 140 & 0.008 & 0.684 & 7 \\
50 & 350 & 0.025 & 0.673 & 7 \\
100 & 700 & 0.098 & 0.670 & 7 \\
200 & 1400 & 0.423 & 0.668 & 7 \\
500 & 3500 & 2.496 & 0.667 & 7 \\
1000 & 7000 & 9.725 & 0.667 & 7 \\
  \end{tabular}%
  \label{tab:iplearner}}
  \qquad\qquad\qquad
\subfloat[10 points]{%
  \begin{tabular}{r|ccc}
$h$ & time (s) & value & nodes \\
    \hline
3 & 0.003 & 0.704 & 7 \\
9 & 0.081 & 0.732 & 23 \\
15 & 0.598 & 0.745 & 47 \\
21 & 2.137 & 0.752 & 79 \\
27 & 6.059 & 0.756 & 119 \\
33 & 13.715 & 0.760 & 167 \\
39 & 30.324 & 0.762 & 223 \\
  \end{tabular}
  \label{tab:iprep}
  }
  \
  \caption{IP results for (repeated) learner}
\end{table}

\begin{figure}\centering
  \begin{tikzpicture}[scale=0.70]
\begin{axis}[
  title={Runtime},
  xlabel=horizon,
  ylabel=time (s),
  legend pos = north west]
            \addplot table [y=IP3, x=h]{figures/rep-time.txt};
\addlegendentry{IP 3 pts}
\addplot table [y=IP4, x=h]{figures/rep-time.txt};
\addlegendentry{IP 4 pts}
\addplot table [y=IP5, x=h]{figures/rep-time.txt};
\addlegendentry{IP 5 pts}
\addplot table [y=IP10, x=h]{figures/rep-time.txt};
\addlegendentry{IP 10 pts}
\addplot table [y=P10, x=h]{figures/rep-time.txt};
\addlegendentry{PBVI 10 pts}
\addplot table [y=P20, x=h]{figures/rep-time.txt};
\addlegendentry{PBVI 20 pts}
\addplot table [y=P50, x=h]{figures/rep-time.txt};
\addlegendentry{PBVI 50 pts}
\addplot table [y=P100, x=h]{figures/rep-time.txt};
\addlegendentry{PBVI 100 pts}
\end{axis}
\end{tikzpicture}
  \hfill
  \begin{tikzpicture}[scale=0.70]
\begin{axis}[
  title={Reachability},
  xlabel=horizon,
  ylabel=value]
\addplot table [y=IP3, x=h]{figures/rep-value.txt};
\addplot table [y=IP4, x=h]{figures/rep-value.txt};
\addplot table [y=IP5, x=h]{figures/rep-value.txt};
\addplot table [y=IP10, x=h]{figures/rep-value.txt};
\addplot table [y=P10, x=h]{figures/rep-value.txt};
\addplot table [y=P20, x=h]{figures/rep-value.txt};
\addplot table [y=P50, x=h]{figures/rep-value.txt};
\addplot table [y=P100, x=h]{figures/rep-value.txt};
\end{axis}
\end{tikzpicture}
  \caption{IP and PBVI results for repeated learner}
  \label{fig:rep}
\end{figure}
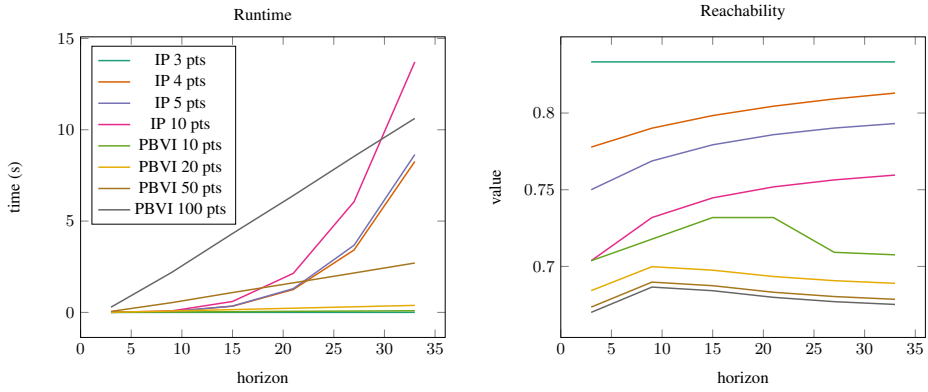

\subsubsection{Robot on a Grid}
Our next case study is a variant of the ever popular Gridworld~\cite{RussellNorvig:2009}. 
The instance that we report on is a $3\times 3$ grid with initial 
state at position $(1,1)$, sink at position $(2,2)$, and target at 
position $(3,3)$.  We want to maximize the probability to reach 
the target from the initial position. 

The robot has (up to) 4 actions available: up, down, left, right.
The actions are probabilistic in the sense that with some probability, 
instead of going forward the robot may end up in the cell to the left 
or to the right. 
We compare two variants: In both variants there is a parameter $p$ 
that describes the total error probability.  In the 1-parameter variant, 
the probability to err left and right is equal to $p/2$; In the 2-parameter 
variant we also include a left-right bias $b$, resulting in probability 
$p*b$ to err left and $p*(1-b)$ to err right. If it is not possible to go left or right, then the other option gets probability $p$. 
For example, the action up in cell $(1,1)$ leads correctly to cell $(2,1)$ with probability $1-p$ and to cell $(1,2)$ on the right with probability $p$ (as no cell is on the left), in both models. The action up in cell $(1,2)$ has the possibility to err left and right, hence shows the difference between the two models.   
See~\cite{ArmingBS17} for a detailed description of robot-on-a-grid models. In that terminology, we use the ``fixed failure" variant.

\subsubsection{Consensus}

The consensus protocol model is the only PARAM benchmark \cite{PARAM:CaseStudies}
that has true nondeterminism in the sense that its policy depends on the 
parameter values. The protocol was introduced by Aspnes and Herlihy~\cite{Aspnes:1990hp}.
The 2-parameter model is exactly the same as the PARAM model, see~\cite{PARAM:CaseStudies} for all details. 
The 1-parameter model depends on a parameter $p$ and is 
obtained by setting $p_1=p$ and $p_2=1-p$, i.e., it is a bias 
parameter with average $1/2$. 
We used $N=K=2$ and the target state is the state in which 
consensus is reached with the preferred value.

Figure~\ref{fig:grid} and Figure~\ref{fig:cons} show the experimental results of the robot on a grid, and the consensus protocol benchmarks, respectively. We only ran PBVI as IP was too slow and in both cases notice that the runtime grows exponentially with the horizon. The grid pMDP model has 10 states and 2 parameters, hence the induced POMDP for 10 parameter points has 1000 states. The consensus protocol model has 273 states and hence the induced POMDP for 5 parameter points has 1365 states.

\begin{figure}\centering
	\begin{tikzpicture}[scale=0.70]
	\begin{axis}[
		title={Runtime},
		xlabel={horizon},
		ylabel={time (s)},
		legend pos = north west]
	\addplot table [y=t2, x=h]{figures/grid.txt};
	\addlegendentry{2 pts}
	\addplot table [y=t5, x=h]{figures/grid.txt};
	\addlegendentry{5 pts}
	\addplot table [y=t7, x=h]{figures/grid.txt};
	\addlegendentry{7 pts}
	\addplot table [y=t10, x=h]{figures/grid.txt};
	\addlegendentry{10 pts}
	\end{axis}
	\end{tikzpicture}
	\hfill
	\begin{tikzpicture}[scale=0.70]
	\begin{axis}[
		xlabel=horizon,
		ylabel=value,
		title={Reachability},
	]
	\addplot table [y=v2, x=h]{figures/grid.txt};
	\addplot table [y=v5, x=h]{figures/grid.txt};
	\addplot table [y=v7, x=h]{figures/grid.txt};
	\addplot table [y=v10, x=h]{figures/grid.txt};
	\end{axis}
	\end{tikzpicture}

  \caption{PBVI results for robot on a grid}
  \label{fig:grid}
\end{figure}
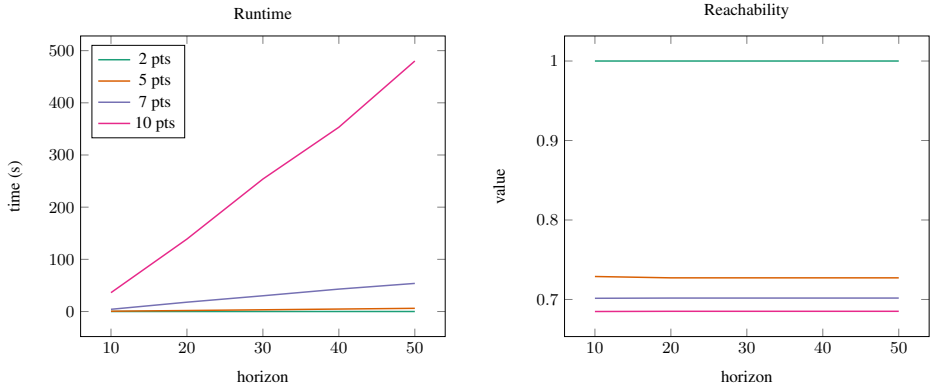

\begin{figure}\centering
	\begin{tikzpicture}[scale=0.70]
	\begin{axis}[
		title={Runtime},
		xlabel={horizon},
		ylabel={time (s)},
		legend pos = north west]
	\addplot table [y=t2, x=h]{figures/consensus.txt};
	\addlegendentry{2 pts}
	\addplot table [y=t3, x=h]{figures/consensus.txt};
	\addlegendentry{3 pts}
	\addplot table [y=t4, x=h]{figures/consensus.txt};
	\addlegendentry{4 pts}
	\addplot table [y=t5, x=h]{figures/consensus.txt};
	\addlegendentry{5 pts}
	\end{axis}
	\end{tikzpicture}
	\hfill
	\begin{tikzpicture}[scale=0.70]
	\begin{axis}[
		xlabel=horizon,
		ylabel=value,
		title={Reachability},
	]
	\addplot table [y=v2, x=h]{figures/consensus.txt};
	\addplot table [y=v3, x=h]{figures/consensus.txt};
	\addplot table [y=v4, x=h]{figures/consensus.txt};
	\addplot table [y=v5, x=h]{figures/consensus.txt};
	\end{axis}
	\end{tikzpicture}
	
	\caption{PBVI results for consensus}
	\label{fig:cons}
\end{figure}
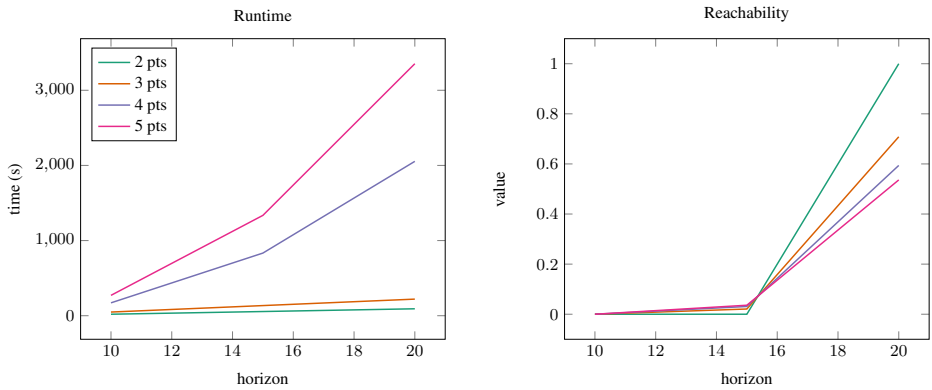

\section{Conclusion}\label{sec:conclusion}

We have presented a way to compute parameter-independent 
strategies that are expectation $\varepsilon$-optimal for pMDP by 
encoding the problem as to compute $\varepsilon$-solutions
in POMDPs.  We have implemented this approach  
using Storm~\cite{DehnertJK017} and AI-Toolbox~\cite{aitoolbox}
and we have evaluated on different case studies. 
Future work will focus on improving the 
efficiency of the current algorithms (for better scalability) by 
taking into account the particular POMDP 
structure resulting from the encoding.

\bibliographystyle{splncs03}
\bibliography{paper.bib}

\begin{thebibliography}{10}
\providecommand{\url}[1]{\texttt{#1}}
\providecommand{\urlprefix}{URL }

\bibitem{ArmingBS17}
Arming, S., Bartocci, E., Sokolova, A.: {SEA-PARAM:} exploring schedulers in
  parametric {M}{D}{P}s. In: Proc. QAPL 2017. {EPTCS}, vol. 250, pp. 25--38
  (2017)

\bibitem{Aspnes:1990hp}
Aspnes, J., Herlihy, M.: {Fast Randomized Consensus Using Shared Memory.} J.
  Algorithms  11(3),  441--461 (1990)

\bibitem{BGB12}
Baier, C., Gr{\"o}{\ss}er, M., Bertrand, N.: Probabilistic $\omega$-automata.
  J. ACM  59(1),  1:1--1:52 (2012)

\bibitem{Baier2008}
Baier, C., Katoen, J.: Principles of Model Checking. {MIT} Press (2008)

\bibitem{BaldiBCCSSS17}
Baldi, M., Bartocci, E., Chiaraluce, F., Cucchiarelli, A., Senigagliesi, L.,
  Spalazzi, L., Spegni, F.: A probabilistic small model theorem to assess
  confidentiality of dispersed cloud storage. In: Proc. QEST 2017. LNCS, vol.
  10503, pp. 123--139 (2017)

\bibitem{aitoolbox}
Bargiacchi, E.: {AI-Toolbox}. \url{https://github.com/Svalorzen/AI-Toolbox/}

\bibitem{BartocciGKRS11}
Bartocci, E., Grosu, R., Katsaros, P., Ramakrishnan, C.R., Smolka, S.A.: Model
  repair for probabilistic systems. In: Proc.~{TACAS} 2011. LNCS, vol. 6605,
  pp. 326--340 (2011)

\bibitem{Beyer:2015ip}
Beyer, D., L{\"o}we, S., Wendler, P.: {Benchmarking and Resource Measurement.}
  In: Proc. SPIN 2015. LNCS, vol. 9232, pp. 160--178 (2015)

\bibitem{Cassandra:1997vz}
Cassandra, A.R., Littman, M.L., Zhang, N.L.: {Incremental Pruning - A Simple,
  Fast, Exact Method for Partially Observable {M}arkov Decision Processes.} In:
  Proc. UAI 1997. pp. 54--61 (1997)

\bibitem{CDH10a}
Chatterjee, K., Doyen, L., Henzinger, T.A.: Qualitative analysis of
  partially-observable {M}arkov decision processes. In: Proc. MFCS 2010. pp.
  258--269 (2010)

\bibitem{DBLP:journals/ai/ChatterjeeC15}
Chatterjee, K., Chmelik, M.: {POMDPs }under probabilistic semantics. Artif.
  Intell.  221,  46--72 (2015)

\bibitem{CCD16}
Chatterjee, K., Chmelik, M., Davies, J.: A symbolic {S}{A}{T}-based algorithm
  for almost-sure reachability with small strategies in {POMDPs}. In: Proc.
  {AAAI} 2016. pp. 3225--3232 (2016)

\bibitem{ChatterjeeCGK16}
Chatterjee, K., Chmelik, M., Gupta, R., Kanodia, A.: Optimal cost almost-sure
  reachability in {P}{O}{M}{D}{P}s. Artif. Intell.  234,  26--48 (2016)

\bibitem{Krish:Randomness-for-free}
Chatterjee, K., Doyen, L., Gimbert, H., Henzinger, T.A.: Randomness for free.
  In: Proc.~MFCS 2010. vol. 6281, pp. 246--257 (2010)

\bibitem{ChenHHKQ013}
Chen, T., Hahn, E.M., Han, T., Kwiatkowska, M.Z., Qu, H., Zhang, L.: Model
  repair for {M}arkov decision processes. In: Proc.~{TASE} 2013. pp. 85--92
  (2013)

\bibitem{Cubuktepe2017}
Cubuktepe, M., Jansen, N., Junges, S., Katoen, J.P., Papusha, I., Poonawala,
  H.A., Topcu, U.: Sequential convex programming for the efficient verification
  of parametric {M}{D}{P}s. In: Proc.~{TACAS} 2017. LNCS, vol. 10206, pp.
  133--150 (2017)

\bibitem{DC05}
Daws, C.: Symbolic and parametric model checking of discrete-time {{M}arkov}
  chains. In: Proc.~{ICTAC} 2004. LNCS, vol. 3407, pp. 280--294 (2005)

\bibitem{DehnertJJCVBKA15}
Dehnert, C., Junges, S., Jansen, N., Corzilius, F., Volk, M., Bruintjes, H.,
  Katoen, J., {\'{A}}brah{\'{a}}m, E.: {PROPhESY}: {A} {PRO}babilistic
  {P}aram{E}ter {SY}nthesis {T}ool. In: Proc.~{CAV} 2015. LNCS, vol. 9206, pp.
  214--231 (2015)

\bibitem{DehnertJK017}
Dehnert, C., Junges, S., Katoen, J., Volk, M.: A storm is coming: {A} modern
  probabilistic model checker. In: Proc. {CAV} 2017. LNCS, vol. 10427, pp.
  592--600 (2017)

\bibitem{HahnHZ11b}
Hahn, E.M., Han, T., Zhang, L.: Probabilistic reachability for parametric
  {M}arkov models. {STTT}  13(1),  3--19 (2011)

\bibitem{HahnHZ11}
Hahn, E.M., Han, T., Zhang, L.: Synthesis for {PCTL} in parametric {M}arkov
  decision processes. In: Proc.~{NFM} 2011. LNCS, vol. 6617, pp. 146--161
  (2011)

\bibitem{PARAM:CaseStudies}
Hahn, E.M., Hermanns, H., Zhang, L., Wachter, B.: {PARAM Case Studies}.
  \url{https://depend.cs.uni-saarland.de/tools/param/casestudies} (2015)

\bibitem{Halmos:MT-Book}
Halmos, P.R.: Measure Theory. Springer (1974)

\bibitem{JansenCVWAKB14}
Jansen, N., Corzilius, F., Volk, M., Wimmer, R., {\'{A}}brah{\'{a}}m, E.,
  Katoen, J., Becker, B.: Accelerating parametric probabilistic verification.
  In: Proc.~{QEST} 2014. LNCS, vol. 8657, pp. 404--420 (2014)

\bibitem{abs-1710-10294}
Junges, S., Jansen, N., Wimmer, R., Quatmann, T., Winterer, L., Katoen, J.,
  Becker, B.: Finite-state controllers of {P}{O}{M}{D}{P}s via parameter
  synthesis. In: Proc. UAI 2018 (2018)

\bibitem{Kaelbling:1998vs}
Kaelbling, L.P., Littman, M.L., Cassandra, A.R.: {Planning and Acting in
  Partially Observable Stochastic Domains.} Artif. Intell.  (1998)

\bibitem{Kreinovich1998bi}
Kreinovich, V., Lakeyev, A., Rohn, J., Kahl, P.: {Computational Complexity and
  Feasibility of Data Processing and Interval Computations}, Applied
  Optimization, vol.~10. Springer (1998)

\bibitem{KwiatkowskaNP11}
Kwiatkowska, M.Z., Norman, G., Parker, D.: {PRISM} 4.0: Verification of
  probabilistic real-time systems. In: Proc.~{CAV} 2011. LNCS, vol. 6806, pp.
  585--591 (2011)

\bibitem{LMST2007}
Lanotte, R., Maggiolo-Schettini, A., Troina, A.: Parametric probabilistic
  transition systems for system design and analysis. Form. Asp. Comput.  19(1),
   93--109 (2007)

\bibitem{LukinaEHBYTSG17}
Lukina, A., Esterle, L., Hirsch, C., Bartocci, E., Yang, J., Tiwari, A.,
  Smolka, S.A., Grosu, R.: {ARES:} adaptive receding-horizon synthesis of
  optimal plans. In: Proc. {TACAS} 2017. LNCS, vol. 10206, pp. 286--302 (2017)

\bibitem{MHC03}
Madani, O., Hanks, S., Condon, A.: On the undecidability of probabilistic
  planning and related stochastic optimization problems. Artif. Intell.
  147(1-2),  5--34 (2003)

\bibitem{Ayala2012}
Medina~Ayala, A.I., Andersson, S.B., Belta, C.: Probabilistic control from
  time-bounded temporal logic specifications in dynamic environments. In:
  Proc.~{ICRA} 2012. pp. 4705--4710. {IEEE} (2012)

\bibitem{Papadimitriou:1987ki}
Papadimitriou, C.H., Tsitsiklis, J.N.: {The Complexity of {M}arkov Decision
  Processes.} Math. Oper. Res.  12(3),  441--450 (1987)

\bibitem{PathakAJTK15}
Pathak, S., {\'{A}}brah{\'{a}}m, E., Jansen, N., Tacchella, A., Katoen, J.: A
  greedy approach for the efficient repair of stochastic models. In:
  Proc.~{NFM} 2015. LNCS, vol. 9058, pp. 295--309 (2015)

\bibitem{PineauGT03}
Pineau, J., Gordon, G.J., Thrun, S.: {Point-based value iteration - An anytime
  algorithm for {P}{O}{M}{D}{P}s.} Proc. IJCAI-03 pp. 1025--1032 (2003)

\bibitem{PolgreenWHA17}
Polgreen, E., Wijesuriya, V.B., Haesaert, S., Abate, A.: Automated experiment
  design for data-efficient verification of parametric {M}arkov decision
  processes. In: Proc. {QEST} 2017. LNCS, vol. 10503, pp. 259--274 (2017)

\bibitem{QuatmannD0JK16}
Quatmann, T., Dehnert, C., Jansen, N., Junges, S., Katoen, J.: Parameter
  synthesis for {M}arkov models: Faster than ever. In: Proc.~{ATVA} 2016. LNCS,
  vol. 9938, pp. 50--67 (2016)

\bibitem{Roy:2005kv}
Roy, N., Gordon, G.J., Thrun, S.: {Finding Approximate {P}{O}{M}{D}{P}
  solutions Through Belief Compression.} J. Artif. Intell. Res.  23,  1--40
  (2005)

\bibitem{RussellNorvig:2009}
Russell, S., Norvig, P.: {Artificial Intelligence: A Modern Approach}. Prentice
  Hall (2009)

\bibitem{Sennott1998}
Sennott, L.I.: Stochastic Dynamic Programming and the Control of Queueing
  Systems. Wiley (1998)

\bibitem{Spaan:2011cs}
Spaan, M.T.J., Vlassis, N.: {Perseus: Randomized Point-based Value Iteration
  for {P}{O}{M}{D}{P}s}. J. Artif. Intell. Res.  24,  195--220 (2011)

\end{thebibliography}

\newpage
\appendix

\section{Proofs} \label{app:proofs}

\begin{proof}[of Lemma~\ref{lem:runs-1-1}]
	It is quite obvious that $f$ is injective and surjective. The isomorphism property follows directly from the observation that $$f(\{x\}\times \Cone(h_x)) = \Cone(f(x,h_x))$$ for any $x \in X$ and $h_x \in \Hist(M(x))$. \qed
\end{proof}

\begin{proof}[of Lemma~\ref{lem:policy-1-1}]
	Let $\pi_X$ be a policy of the pMDP $M$ and let $M'$ be the induced POMDP. Let $h_1, h_2 \in \Hist'$ be two histories with $O(h_1) = O(h_2)$. Then $f^{-1}(h_1)$ and $f^{-1}(h_2)$ are in $\Hist_X$ and they only differ in the first coordinate, i.e., the parameter value. Hence $\pi'(h_1) = \pi_X(f^{-1}(h_1)) =  \pi_X(f^{-1}(h_2)) = \pi'(h_2)$ showing that $\Phi$ is well defined. 
	
	Similarly, let $\pi'$ be a policy of $M'$ and let $h_x, h_y \in \Hist_X$ be such that $h_x = (x,h)$ and $h_y = (y,h)$, i.e., for any pMDP policy $\pi$, $\pi(h_x) = \pi(h_y)$. We have $O(f(h_x)) = h = O(f(h_y))$ and hence
	$\pi'(f(h_x)) = \pi'(f(h_y))$ yielding that indeed $\pi_X(h_x) = \pi_X(h_y)$. This shows that $\Phi^{-1}$ is well defined too.
	
	The rest follows from the observation that $\Phi^{-1}$ is indeed an inverse of $\Phi$ as 
	$$\Phi^{-1}(\Phi(\pi_X)) = \Phi^{-1}(\pi_X \circ f^{-1}) = (\pi_X \circ f^{-1}) \circ f = \pi_X$$
	and 
	$$\Phi(\Phi^{-1}(\pi')) = \Phi(\pi'\circ f) = (\pi' \circ f)\circ f^{-1} = \pi'.$$\qed
\end{proof}

\begin{proof}[of Lemma~\ref{lem:measure-eq}]
    It suffices to show that the measures coincide on all generators, i.e.
    \[
        \Pp{}^{M}_{\pi_X,i,p}(\{x\}\times\Cone(h))=\Pp{}^{M'}_{\pi',i'}(\Cone(f(x,h)))
    \] for all $x\in X$ and $h\in \Hist(M(x))$.
We prove this by induction on $n$ in $h=s_0, a_0, s_1, a_1\cdots s_n$. For $n=0$, $h = s_0$ and we have
\begin{eqnarray*}
	\Pp{}^{M}_{\pi_X,i,p}(\{x\} \times \Cone(s_0)) & = & p(x) \cdot \Pp{}^{M(x)}_{\pi_x,i}(\Cone(s_0))\\
	& = & p(x)\cdot i(s_0)\\
	& = & i'(s_0,x) \\
    & = &   \Pp{}^{M'}_{\pi',i'}(\Cone(f(x,s_0)). 
\end{eqnarray*}
\sloppypar{For $h = s_0, a_0, s_1, a_1 \dots s_n, a_{n}, s_{n+1}$, assuming the property holds for $w = s_0, a_0, s_1, a_1, \dots s_n$, we get } 
    \begin{eqnarray*}
 & & \hspace*{-1.3cm}       \Pp{}^{M}_{\pi_X,i,p}(\{x\} \times \Cone(h)) \\
\qquad\qquad & = &       p(x) \cdot \Pp{}^{M(x)}_{\pi_x,i}(\Cone(h))\\
\qquad\qquad & = &         p(x) \cdot \Pp{}^{M(x)}_{\pi_x,i}(\Cone(w)) \cdot \pi_x(w)(a_n) \cdot T(s_n,a_n)(x)(s_{n+1}) \\
\qquad\qquad & = &         \Pp{}^{M}_{\pi_X,i,p}(\{x\}\times \Cone(w)) \cdot \pi_X(x,w)(a_n) \cdot T(s_n,a_n)(x)(s_{n+1}) \cdot \delta_x(x) \\
\qquad\qquad & \stackrel{(IH)}{=} &         \Pp{}^{M'}_{\pi',i'}(\Cone(f(x,w)) \cdot \pi'(f(x,w))(a_n) \cdot T'((s_n,x),a_n)(s_{n+1},x) \\
\qquad\qquad & = &         \Pp{}^{M'}_{\pi',i'}(\Cone(f(x,h))) . \hspace*{6.5cm}\qed
    \end{eqnarray*} 
\end{proof}

\begin{proof}[of Theorem~\ref{thm:correspondence}]
    From Lemma~\ref{lem:runs-1-1}, $f$ is measurable. Using Theorem~\ref{thm:halmos} at the equality marked by $(*)$ and Lemma~\ref{lem:measure-eq} at the equality marked by $(**)$ we get:
    \begin{eqnarray*}
        \Exp^M_{\pi_X,i,p}(r) & = & \int_{\Runs_X} r \, d\Pp{}^M_{\pi_X,i,p}\\
        & \stackrel{(*)}{=} & \int_{\Runs'} (r \circ f^{-1}) \, d\left(\Pp{}^{M}_{\pi_X,i, p} \circ f^{-1}\right)\\
        & \stackrel{(**)}{=} & \int_{\Runs'} r' \, d\Pp{}^{M'}_{\pi',i'}\\
        & = & \Exp^{M'}_{\pi',i'}(r').   \hspace*{8.4cm}\qed 
   \end{eqnarray*}
\end{proof}

\section{AI Toolbox Algorithms and Policy Representation}
\label{app:AI-toolbox}

POMDP algorithms come in two flavors: offline and online.
Online algorithms compute an optimal action for a given belief, whereas
offline algorithms compute a policy for every belief.
Intuitively, an online algorithm \emph{is} a policy and an offline algorithm \emph{returns} a policy.
In this work we focus on offline algorithms only, as we wish to evaluate the policy produced. We leave experimenting with online algorithms for future work. Moreover, offline and online algorithms are somewhat interchangeable: one can use an offline algorithm online, and one can expand an online algorithm (up to some finite horizon) to construct a policy.

The output of an offline algorithm is a \emph{policy graph} together with a \emph{value function} for each node of the graph.
\begin{definition}[Policy graph, value function]
	A \emph{policy graph} for a POMDP $M = (S, A, T, i, \Omega, O)$ is a tuple $(V,E, L_n, L_e)$ where $(V,E)$ is a directed graph with node labels from $A$ given by $L_n\colon V \to A$ and edge labels from $\Omega$ given by $L_e\colon E \to \Omega$, such that for every $v\in V$ and every $o\in\Omega$ precisely one outgoing edge from $v$ is labelled $o$. The \emph{value function} is a family of functions  $f_v\colon S \to \Reals$ for each $v \in V$. 
\end{definition}
Each node $v$ corresponds to a policy in the following way: first take the action associated with $v$ and then follow the edge corresponding to the observation to a new node $v'$, and iterate.
The value function $f_v(s)$ for a node $v$ and a state $s$ gives the expected return of the policy $v$ on initial state $s$.

Since POMDPs are in general intractable~\cite{Papadimitriou:1987ki}, most research focusses on finding approximate solutions and bounds, and there are only two exact algorithms implemented in AI-Toolbox: \emph{Witness}~\cite{Kaelbling:1998vs} and \emph{Incremental Pruning} (IP)~\cite{Cassandra:1997vz}.
In our preliminary evaluations IP outperformed Witness -- an observation consistent with \cite{Cassandra:1997vz}. We thus selected IP to compute exact solutions where possible.
As expected for an exact algorithm, we found IP to only work on relatively small examples.

The remaining offline algorithms offered by AI-Toolbox are: AMDP~\cite{Roy:2005kv}, PERSEUS~\cite{Spaan:2011cs} and PBVI.
Both PBVI and PERSEUS are examples of \emph{point-based} algorithms,
an important class of POMDP algorithms that consider only some small (usually sampled) set of points in the belief space.
These algorithms represent the state of the art in general POMDP solvers %
and share the nice property of providing lower bounds.
In contrast, we are not aware of any guaranteed (error-) bounds for AMDP.
PERSEUS only works in the discounted case and is thus not suitable for reachability objectives.

\end{document}